\documentclass[12pt,preprint]{aastex}
\usepackage{amsmath}

\shorttitle{Disrupted asteroid P/2016 G1}
\shortauthors{Moreno et al.}

\begin{document}


\title{Early evolution of disrupted asteroid P/2016 G1 (PANSTARRS)}


\author{F. Moreno\affil{Instituto de Astrof\'\i sica de Andaluc\'\i a, CSIC,
  Glorieta de la Astronom\'\i a s/n, 18008 Granada, Spain}
\email{fernando@iaa.es}}

\author{
J. Licandro\affil{Instituto de Astrof\'\i sica de Canarias,
  c/V\'{\i}a 
L\'actea s/n, 38200 La Laguna, Tenerife, Spain, 
\and 
 Departamento de Astrof\'{\i}sica, Universidad de
  La Laguna (ULL), E-38205 La Laguna, Tenerife, Spain}}  

\author{
A. Cabrera-Lavers\affil{Instituto de Astrof\'\i sica de Canarias,
  c/V\'{\i}a 
L\'actea s/n, 38200 La Laguna, Tenerife, Spain, 
\and 
 Departamento de Astrof\'{\i}sica, Universidad de
  La Laguna (ULL), E-38205 La Laguna, Tenerife, Spain, 
\and 
Gran Telescopio Canarias (GTC), E-38712, Bre\~na Baja, La Palma, Spain}}

\and

\author{F.J. Pozuelos\affil{Instituto de Astrof\'\i sica de Andaluc\'\i a, CSIC,
  Glorieta de la Astronom\'\i a s/n, 18008 Granada, Spain} }


\begin{abstract}

We present deep imaging observations   
of activated asteroid P/2016 G1 (PANSTARRS) using the 10.4m Gran Telescopio Canarias (GTC) 
from late April to early June 2016. The images are best interpreted as the
result of a relatively short-duration event with onset about
$\mathop{350}_{-30}^{+10}$  days  
before perihelion (i.e., around 10th February, 2016), starting sharply
and decreasing with a $\mathop{24}_{-7}^{+10}$ days
(Half-width at half-maximum, HWHM). The results of the modeling 
imply the emission of $\sim$1.7$\times$10$^7$ kg of dust, if composed
of particles of 1 micrometer to 1 cm in radius, distributed
following a power-law of index --3, and having a geometric albedo of 0.15.  A detailed fitting of a
conspicuous westward feature in the head of the comet-like object  
indicates that a significant fraction of the dust was ejected along a
privileged direction right at the beginning of the event, 
which suggests that the parent body has possibly 
suffered an impact followed by a partial or total disruption. From the
limiting magnitude reachable with the instrumental setup, and assuming
a geometric albedo of 0.15 for the parent body, an upper limit for
the size of possible fragment debris of $\sim$50 m in radius is derived.

\end{abstract}


\section{Introduction}
P/2016 G1 (PANSTARRS)  (hereafter P/2016 G1 for short) 
was discovered by R. Weryk and R. J. Wainscoat 
on CCD images acquired on 2016 April 1 UT with the 
1.8-m Pan-STARRS1 telescope \citep{Weryk16}. From the derived orbital elements
($a$=2.853 AU, $e$=0.21, $i$=10.97$^\circ$), its Tisserand parameter
respect to Jupiter \citep{Kresak82} can be calculated as $T_J$=3.38,
so that the object belongs dynamically to the main asteroid belt. 
The discovery images revealed however a cometary appearance showing clear
evidence of a tail extending for approximately 20\arcsec, 
and a central condensation broader than field stars \citep{Weryk16}. Since the
discovery of the object   
133P/Elst-Pizarro in 1996 \citep[see e.g.][and references
  therein]{Hsieh04}, 
about twenty objects of this class have been
discovered, whose activity triggering mechanisms have been proposed to
range from impact-induced to rotational disruption, while the 
activity has been found to last from a few days or less to a few months. In
this latter case, sublimation-driven of volatile ices has been invoked
as the most likely mechanism of dust production, although gaseous
emissions lines have remained undetected to date. For a review of the
different objects discovered so far, their orbital stability,  and
their activation mechanisms, we refer to \cite{Jewitt15}.

In this paper we report observations of P/2016 G1 acquired with the
10.4m GTC, and present models of the dust tail brightness 
evolution from late April to early June 2016. We provide the onset
time, the total
dust loss, and the duration of the activity, and attempt to identify
which physical mechanism is involved in its activation. 

\section{Observations and data reduction}

Observations of P/2016 G1 were scheduled immediately after the discovery alert,
within our long-term GTC program of activated asteroids
observations. CCD images  
of P/2016 G1 have been obtained under photometric and excellent seeing 
conditions on 
the nights of 20 April, 28 May, and 8 June 2016. The
images were  obtained using a Sloan $r^\prime$ filter in the
Optical System for Image and Low Resolution Integrated Spectroscopy
(OSIRIS) camera-spectrograph \citep{Cepa00,Cepa10} at the GTC. The
plate scale was 0.254$\arcsec$ px$^{-1}$. The images were bias subtracted,
flat-fielded, and their flux was calibrated using standard stars from
 \cite{Smith02}. Those stars are observed with a slight
 defocus applied to the telescope in order to get high S/N in one
 second exposures, needed to avoid shutter non-uniformity effects in
 the photometry (that is accurate to 1\% level). A 
median stack image was produced for each night of observation from the available
frames. These images were converted from mag arcsec$^{-2}$ 
to solar disk intensity units (the output of our Monte Carlo dust tail
code) by setting $r_\odot^\prime$=--26.95,
obtained assuming $V_{\odot}$=--26.75 and $(B-V)_{\odot}$=0.65
\citep{Cox00}, and the photometric relations from
\cite{Fukugita96}. The log of the observations is presented in Table
1. This table includes the date of the observations (in UT and
in days to perihelion), the heliocentric ($R$) and geocentric ($\Delta$)
distances, the solar phase angle ($\alpha$), the position angle of the Sun
to comet radius vector (PsAng), and the angle between the Earth and
the asteroid orbital plane (PlAng). The reduced images are shown in
Figure 1, in the conventional North-up, East-to-the-left orientation,
and showing the directions of the Sun and orbital motion of the
object. 

The elevation of the
Earth above the asteroid orbital plane ranges from nearly edge-on
view (PlAng=--0.9$^\circ$ on April 21) to PlAng=--6.1$^\circ$ on June 8, allowing
different viewing angles of the tail for an adequate analysis in terms
of dust tail 
models. The conspicuous appearance of the object, with the lack of a central
 condensation (nucleus), and an inverted-C-shaped head with some
westward extension that becomes 
apparent as PlAng becomes larger and larger, 
immediately suggest a likely disruptive phenomenon
as the cause of the observed activity. Dust motion between the first
and last observation can also contribute to the morphology changes.  
In those images, no small condensations 
that could be attributed to small fragments, \citep[as detected for
P/2013 R3 or P/2012 F5, see][]{Jewitt14,Drahus15}, are seen, however. 
The limiting magnitude with OSIRIS Sloan r$^\prime$ filter for 
getting a signal-to-noise ratio S/N=3, assuming a dark background, a 
seeing disk of FWHM=1.0$\arcsec$, and an airmass of 1.2, would be, 
for the total exposure 
time of 1080 s of the night of 28 May, 2016, of
r$^\prime$=25.8 (see http://www.gtc.iac.es/instruments/osiris/). 
For the observational geometric 
conditions of that night (see Table 1), a spherical
body of 47 m in radius, characterized by a  
geometric albedo of 0.15, and a linear phase coefficient of 0.03 mag 
deg$^{-1}$, would have this magnitude. Hence, no parent body
or fragments larger than that size would have been produced as a
consequence of the asteroid activation. This is actually an optimistic
size limit, however, since the limiting magnitude refers to a dark background,
and not to a source located within a bright background coma. 

\section{The Model}

To perform a theoretical interpretation of the obtained images in
terms of the dust physical parameters, we used our Monte Carlo dust 
tail code, which has been used previously on several works on
activated asteroids and comets, including comet
67P/Churyumov-Gerasimenko, the Rosetta 
target \citep[e.g.,][]{Moreno16a}. This model computes the dust tail
brightness of a comet or activated asteroid by adding up the
contribution to the brightness of each particle ejected from the
parent nucleus. The particles, after leaving the object's surface, are
ejected to space experiencing the solar gravity and radiation
pressure. The nucleus gravity force is neglected, a valid approximation for
small-sized objects. Then, the trajectories of the particles become
Keplerian, having orbital elements which depend on their physical
properties and ejection velocities \citep[e.g.][]{Fulle89}. In order
to build up an usable representation of the individual images with the 
Monte Carlo procedure, we usually launch from 2$\times$10$^6$ to 10$^7$
particles. For a given set of dust parameters, three synthetic images
corresponding to the observational parameters of table 1 
are generated using separate Monte Carlo runs for each image.

The  ratio of radiation
pressure to the gravity forces exerted on each particle is given by
the parameter $\beta =C_{pr}Q_{pr}/(2\rho r)$,  where
$C_{pr}$=1.19$\times$ 10$^{-3}$ kg m$^{-2}$, $Q_{pr}$ is the radiation
pressure coefficient, and $\rho$ is the particle density. $Q_{pr}$ 
is taken as 1, as it converges to that value for
absorbing particles of radius $r \gtrsim$1 $\mu$m 
\citep[see e.g.][their Figure 5]{Moreno12}. 

To make the problem tractable, a number of simplifying assumptions 
on the dust physical parameters must be made. Thus, the
particle density is taken as 1000 kg 
m$^{-3}$, and the geometric albedo is set to $p_v$=0.15, a typical
value for C-type asteroids. The assumption of a lower albedo would
imply an increase in 
the derived loss rates to fit the observed brightness. For the 
particle phase function correction, we use a linear phase coefficient 
of 0.03 mag deg$^{-1}$, which is in the range of comet 
dust particles in the 
1$^\circ \le \alpha \le$ 30$^\circ$ phase angle domain  \citep{MeechJewitt87}. A
broad size distribution is assumed, with minimum
and maximum particle radii set to 1 $\mu$m and 1 cm, respectively, and 
following a power-law function of index $\kappa$=--3. This index was
found appropriate after repeated experimentation with the code, and it
is within the range of previous estimates of the
size distribution of particles ejected from activated asteroids and
comets. 

We assume isotropic ejection of the particles, which will
provide a first order description of the dust model parameters,
although some of them such as the dust size distribution, the
ejected dust mass, and the timing of event, can be estimated using such
approximation.

The ejection velocity of the particles will depend on the activation mechanism
involved, which, in principle, is unknown. However, the object is
located in the inner region of the main belt, having a small 
semi-major axis, suggesting that ice sublimation is unlikely the
driver of the activity. Recent results on 
activated asteroid P/2015 X6 \citep{Moreno16b} reveal that a random
function for the velocities of the form $v = v_1 + \zeta v_2$, 
where $\zeta$ is a random number in the $[0,1]$ interval, and $v_1$
and $v_2$ are fitting parameters, produced adequate results. Hence, to
limit the number of free parameters, we assume that velocity law.

For the dust loss rate as a function of time, we adopt a
 half-Gaussian function whose
maximum is the peak dust-loss rate ($\dot{M}_0$),  located
at the event onset ($t_0$). The half-width at
half-maximum of the Gaussian (HWHM) is a measure of the effective time
span of the event. 

Summarizing, we have selected five fitting parameters for the
isotropic model: the 
two dust ejection velocity parameters ($v_1$ and $v_2$) and the three
parameters associated to the dust loss rate function ($\dot{M}_0$,
$t_0$, and HWHM). The model analysis, aimed at finding the best-fit
set of parameters, is conducted by the downhill
simplex method \citep{Nelder65}, using the FORTRAN implementation
described in \cite{Press92}.  A preliminary, zero-th order analysis, 
of the images is first
performed by building a syndyne-synchrone map \citep{Finson68} for
each observing date. As an example, the synchrone
map for the 28 May 2016 image is displayed in Figure 2. It is clear from the plot
that the time span the asteroid 
has been active should be between around --300 to --360 days to
perihelion, and its duration cannot be much longer than the time
difference of $\sim$60 days, otherwise the dust cloud would have been much more fan
broadly. Hence, parameter $t_0$ was 
assumed in the five-dimensional starting simplex to vary between these
limits. In addition, it is also clear from the synchrone plot that a
very short emission scenario (say, less than a day) can also be ruled
out owing to the observed tail width.  

All the other parameters were assumed to vary broadly between
physically reasonable limits. The fits are characterized by the
parameter $\chi=\sum \sigma_i$, where the summation is extended
to the three images under consideration, and     
$ \sigma_i=\sqrt{(\sum[\log(I_{obs}(i))-\log(I_{fit}(i))]^2/N(i))}$, where
$I_{obs}(i)$  and $I_{fit}(i)$ are the observed and modeled tail
brightness, and $N(i)$ is the number of pixels of image $i$. We work
on the logarithm of intensities in order to give a much more similar weight to
the outermost image isophotes and the innermost ones than would be
done using simply intensities, owing to the brightness
gradient toward the innermost portion of the object.

\section{Results and discussion}

Using the procedure described in the previous section, we found the
following best-fit parameters: $\dot{M}_0$=7.6 kg s$^{-1}$,
$t_0$=--350 days, HWHM=24 days, $v_1$=0.015 m s$^{-1}$  and $v_2$=0.122
m s$^{-1}$. The total dust mass ejected was 1.7$\times$10$^7$ kg, 
all this mass being emitted before the first observation of April 21, 2016. 

The best-fit images are compared to the observed ones in Figure 3. This best-fit
model has $\chi$=0.109. As in our analysis of activated asteroid
P/2015 X6 \citep{Moreno16b},
acceptable solutions are considered only when $\chi \le$0.15. These
limiting values of $\chi$ provide lower and upper limits to the derived
best-fit parameters. Thus, for the timing parameters $t_0$ and HWHM,
we obtain $t_0$=--350$^{+10}_{-30}$ days, and HWHM=24$^{+10}_{-7}$
days. This implies a relatively short-duration event, that, in
principle,  can be associated to a wide range of phenomena. The particle
velocities found are very small, ranging from 0.015 to 0.14 m
s$^{-1}$. The mean value of these velocities ($\sim$ 0.08 m s$^{-1}$)
is comparable to the escape velocity an object of 35 m in radius and
3000 kg m$^{-3}$ in density. However, as stated before, the minimum 
detectable object radius is $\sim$50 m. This implies that a parent nucleus,
or possible fragments, smaller than that size would have remained
unobserved. Higher resolution and better sensitivity images (such as
provided by the Hubble Space Telescope) are clearly needed to
search for fragments. 

Regarding the particle size limits, we must note that the upper limit assumed of $r_{max}$=1 cm is
constraining the total mass ejected to $\sim$1.7$\times$10$^7$
kg. However, if this upper limit is increased, the amount of
mass ejected would be larger accordingly. For instance, increasing $r_{max}$
to 10 cm, we can obtain fits of comparable quality to those shown in Figure 3,
just by varying the power index of the size distribution from
$\kappa$=--3.0 to $\kappa$=--3.2, and the peak mass loss rate from $\dot{M}_0$=7.6 kg
s$^{-1}$ to $\dot{M}_0$=32 kg s$^{-1}$. In consequence, the 
amount of dust mass calculated from our standard isotropic model would be
actually a lower limit to the total mass emitted, if particles larger
than 1 cm in radius were ejected.

In order to gain more insight into the possible activation
mechanism(s), we note that while the inverted-C shape
feature observed on May 29 and June 8 images are approximately
mimicked with the isotropic model, the westward extension is, as
expected, not reproduced in the simulated images, implying some
asymmetry in the ejection pattern. After some
experimentation with the code, we found that if right at the beginning
of the event all the ejected material is directed towards a specific
direction and during a very short time interval, instead of being emitted 
isotropically, we could then simulate that brightness
feature. To accomplish this task, we introduce a cometocentric reference system of unit vectors
($u_r$, $u_\theta$, $u_z$), where $u_r$ 
points away from the Sun, $u_\theta$ is perpendicular to $u_r$,
located on the orbital plane, in the
sense opposite to the comet motion, and $u_z$ is perpendicular to the
orbit plane. We found that in order to simulate the westward
brightness feature, the direction of ejection must satisfy u$_r
\sim$1, i.e., pointing away from the Sun. Figure 4 shows
the effect of ejection around a specific direction given by $u_r
\sim$0.98, $u_\theta \sim$0.18, $u_z \sim$0.08 on the May and June
images (modified isotropic model), compared with the nominal isotropic
model. The duration of this early dust
mass ejection was set to 9 hours, but the feature could be equally
simulated assuming a shorter time interval, as long as the total mass 
ejected in that direction keeps constant.  This could be interpreted as the
result of an impact whose ejecta is directed along that
direction. Although such event should have likely generated an
ejection cone with an aperture $\sim$40$^\circ$, we cannot, however, describe 
the direction pattern more accurately owing to the limited spatial
resolution of the measurements and the viewing angles restriction 
inherent to Earth-based observations. The dust mass ejected in that
privileged direction would be $\sim$2.4$\times$10$^5$ kg. We note that, although
a fraction of the ejected material would have probably traveled at
much higher velocity, in a collision most of the material is actually ejected
at lower velocities \citep{Housen11}.  The impact would have induced a
partial destruction of the asteroid, with dust grains being emitted to
space nearly isotropically while the body is being torn apart. The presence of small
fragments that could be generated in the disruption process can only be assessed with more 
sensitive instrumentation. 
 
\section{Conclusions}  

From the GTC imaging data and the Monte Carlo dust tail modeling of
the activated asteroid P/2016 G1, we arrived to the following conclusions:  

1) Asteroid P/2016 G1 was activated 350$^{+10}_{-30}$ days before
perihelion, i.e., around 10th February 2016. The activity
had a duration of 24$^{+10}_{-30}$ days
(HWHM), so that no dust has been produced since our first observation
of April 21, 2016. The total dust mass emitted was at least $\sim$2$\times$10$^7$ kg,
with a maximum level of activity of $\sim$8 kg s$^{-1}$. These
parameters were estimated assuming a power-law size distribution of
particles between 1 $\mu$m and 1 cm, with power index of
$\kappa$=--3.0, geometric albedo of 0.15, 
and being emitted isotropically from an otherwise
undetected nucleus. The calculated peak and total dust mass are lower
limits, as if larger values for the maximum particle size were assumed, these
quantities would increase. In addition, if different values
for the geometric albedo and/or for the density of the particles were assumed,
these quantities would also change accordingly.

2) While the inverted-C feature which is apparent in the out-of-plane
images of May 29 and June 8, 2016 is approximately mimicked by 
the isotropic ejection model, a westward brightness feature cannot be
reproduced with that model. However, if some dust mass is ejected from
a specified direction right at the time of activation, which turns out
to be approximately along the Sun-to-asteroid vector, that feature
becomes apparent 
in the simulations. We speculate that this dust ejection could be
associated to an impact, and that the subsequent modeled activity is due to
the asteroid partial or total disruption. The impact itself had produced
the ejection of some 2.4$\times$10$^5$ kg of dust. 

3) The inferred ejection velocities of the dust particles 
are very small, in the range of 0.015 to 
0.14 m s$^{-1}$, with an average value of $\sim$0.08 m s$^{-1}$,
corresponding to the escape velocity of an object of 35 m radius and
3000 kg m$^{-3}$ density. An object of that size would have been remained
well below the detection limit of the images acquired, so that we cannot
assure whether fragments of that size or smaller could exist in the
vicinity of the dust cloud. Deeper imaging of the object is clearly 
needed to assess this fact and to determine the fragment  
dynamics.

\acknowledgments

We are very grateful to an anonymous referee for his/her appropriate
and very constructive comments that helped to improve the paper considerably. 

This article is based on observations made with the Gran Telescopio
Canarias, installed in the Spanish Observatorio del Roque de los
Muchachos of the Instituto de Astrof\'\i sica de Canarias, in the island 
of La Palma. 

This work was supported by contracts AYA2015-67152-R  and
AYA2015-71975-REDT from the Spanish
Ministerio de Econom\'\i a y Competitividad. J. Licandro gratefully
acknowledges support from contract AYA2015-67772-R.

\clearpage

\begin{figure}[ht]
\centerline{\includegraphics[scale=0.8,angle=-90]{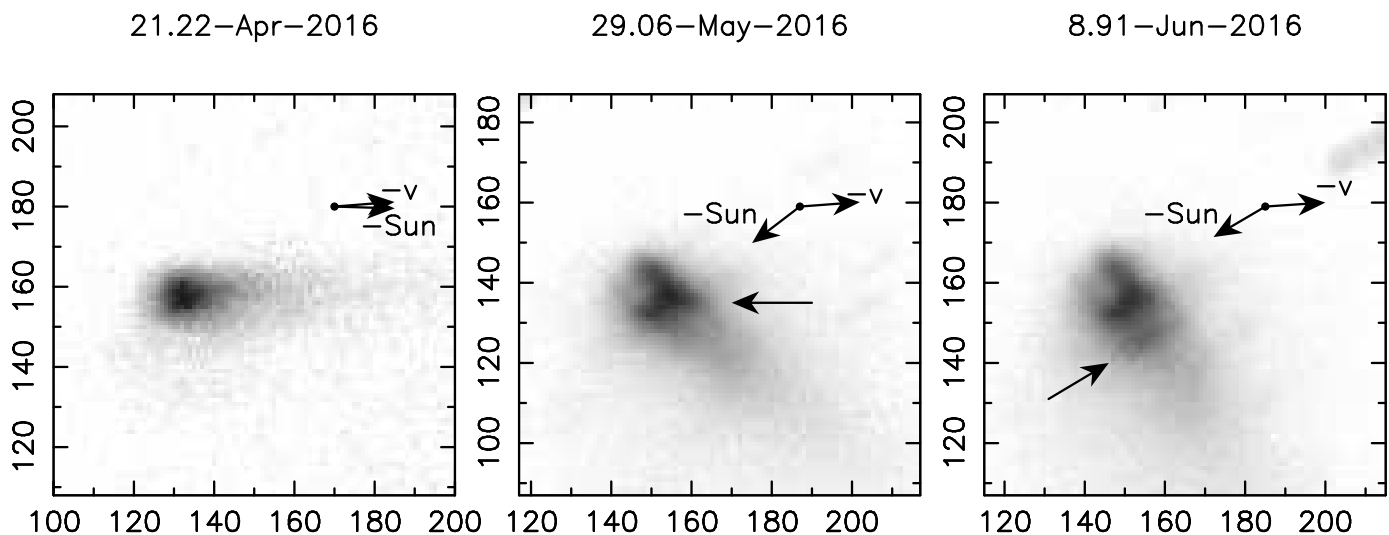}}
\caption{Median stack images of P/2016 G1 obtained with the OSIRIS
  instrument of the 10.4m GTC through a 
Sloan $r^\prime$ filter, at the indicated dates.   North is up, East
to the left.   The 
directions opposite to Sun and the negative of the orbital velocity motion 
are shown. The arrow in the middle of central panel indicates the
westward feature that emerges from the inverted C-shaped mentioned in
the text. The dimensions of the panels (from left to right, in km
projected on the sky at 
the asteroid distance) are 
27930$\times$27930, 26305$\times$26305, and 27025$\times$27025. 
The images are stretched linearly in brightness, 
with maximum intensity
levels, from left to right, of 8$\times$10$^{-14}$, 5$\times$10$^{-14}$, and
4$\times$10$^{-14}$ solar disk intensity units. Faint
trailed stars are  
apparent near the head of the object, perpendicular to the tail, in
the 8.91 June 2016 image, the brightest one being indicated by an
arrow.}
   \label{fig1}
\end{figure}
\clearpage

\begin{figure}[ht]
\centerline{\includegraphics[scale=0.6,angle=-90]{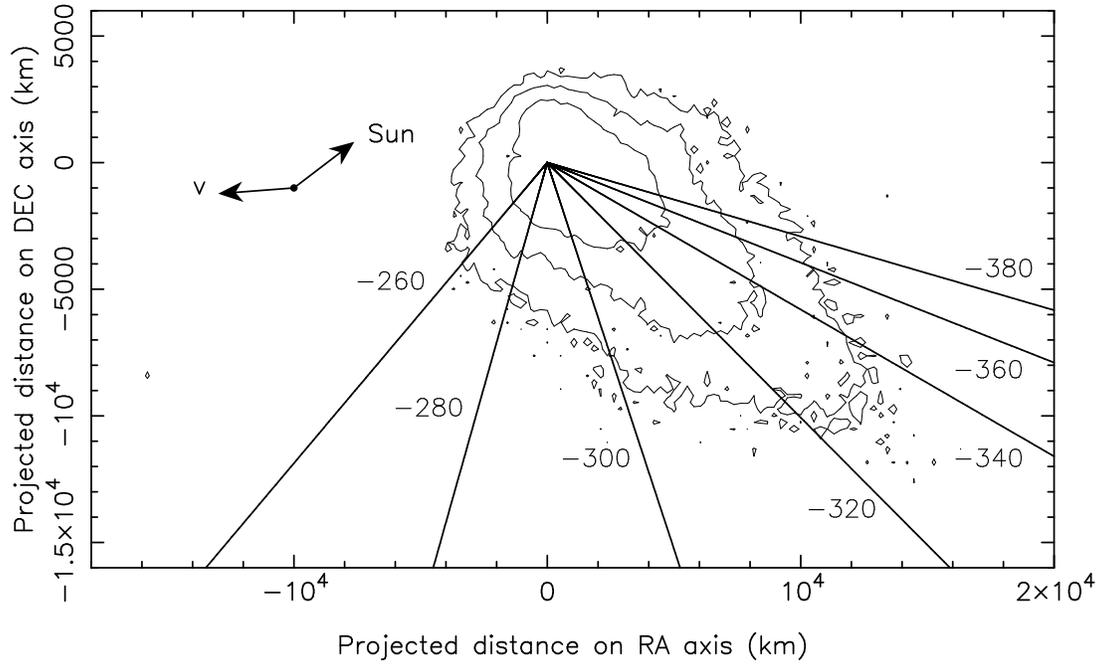}}
  \caption{Synchrone map corresponding to the 29.06 May 2016 image 
    overplotted on the contour map (thin solid lines). Synchrones
    (thick solid lines) correspond to --380
    to --260 days to perihelion, in steps of 20 days, as labeled. The directions to the Sun
    and the orbital velocity motion of the asteroid are indicated.}
  \label{fig2}
\end{figure}

\clearpage

\begin{figure}[ht]
\centerline{\includegraphics[scale=0.7,angle=-90]{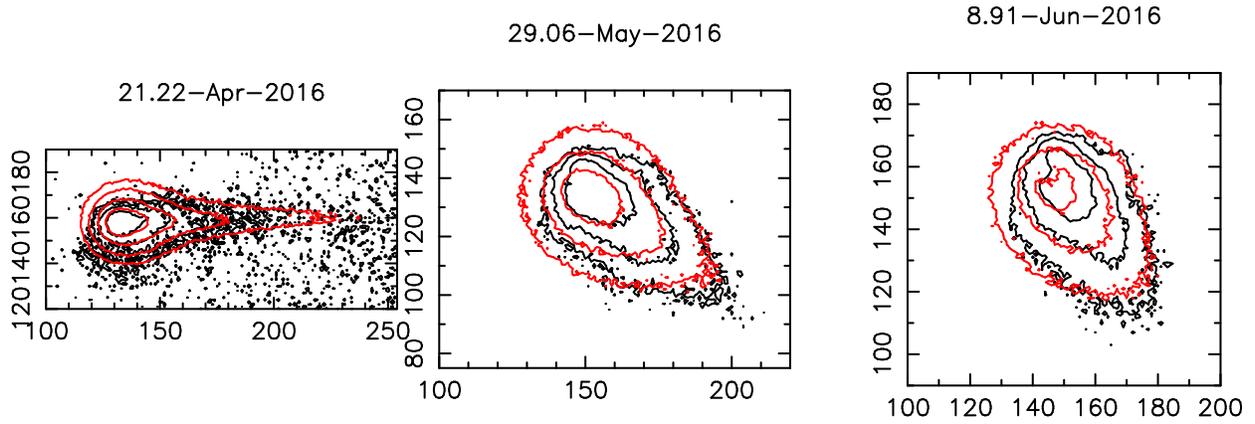}}
  \caption{Measured (black) and modeled (red) isophotes for the three
    images analyzed. Innermost
    isophote levels are 4$\times$10$^{-14}$ (left panel) and
    2$\times$10$^{-14}$ (center and right panels) solar disk intensity
    units, and decrease in factors of two outwards.}
  \label{fig3}
\end{figure}

\clearpage

\begin{figure}[ht]
\centerline{\includegraphics[scale=0.7,angle=-90]{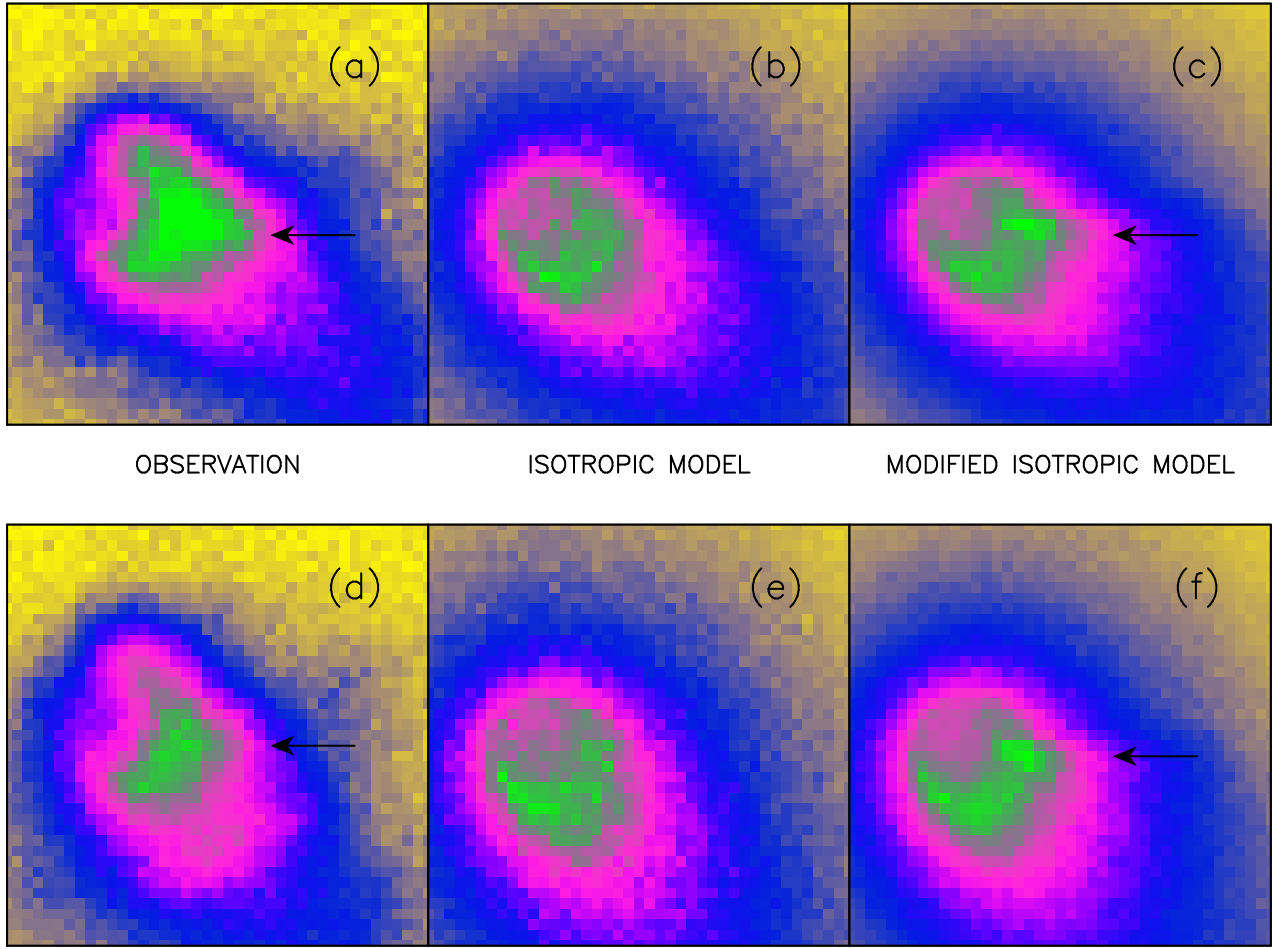}}
\caption{Observed and simulated images in the vicinity of the head region on May
  29.06 (upper panels, a, b, and c) and June 8.91 (lower panels, d, e,
  and f). The observations are on panels (a) and (d). The isotropic
  ejection model images are on panels (b) and (e), and the modified isotropic
  model including early dust ejection in the direction of unit vector
  (0.98,0.18,0.08) in the cometocentric frame are on panels (c) and
  (f). The arrows indicate the location of the westward extension
  described in the text, which appears in the observations and in the
  modified isotropic model, but not in the isotropic model. The upper 
  panels have dimensions of
  10522$\times$10522 km, and the lower panels 10810$\times$10810 km.}
\label{fig4}
\end{figure}

\clearpage

\begin{deluxetable}{cccccccccc}
\rotate
\tablewidth{0pt}
\tablecaption{Log of the observations}
\tablehead{
\colhead{UT (Start) } & \colhead{Days to} &  \colhead{No. of} &
\colhead{Total} & \colhead{Seeing} &
\colhead{R} & \colhead{$\Delta$} &  \colhead{$\alpha$}&  \colhead{PsAng}
& \colhead{PlAng} \\
\colhead{YYYY/MM/DD HH:MM} & \colhead{perihelion} &  \colhead{images}
& \colhead{exp. time (s)} &  \colhead{FWHM(\arcsec)} &
\colhead{(AU)} & \colhead{(AU)} &  \colhead{($^\circ$)}&  \colhead{($^\circ$)}
& \colhead{($^\circ$)} \\
}
\startdata
2016/04/21 05:18 & --297 & 20 & 600 &  1.0 & 2.473 & 1.516 &  9.13 &
268.8 & --0.909 \\
2016/05/29 01:24 & --242 & 6 & 1080 &  0.9 & 2.386 & 1.428 & 10.27 &
127.4 & --5.365 \\
2016/06/08 21:45 & --231 & 5 & 900 &  0.7 & 2.362 & 1.467 & 14.83 &
120.5 & --6.128 \\
\enddata
\end{deluxetable}
\end{document}